\documentclass{llncs} 

\pdfoutput=1

\newif\ifpaper\papertrue

\usepackage[latin1]{inputenc}
\usepackage{float}
\usepackage{alltt}
\usepackage{xspace}
\usepackage{epsfig}
\usepackage{wrapfig}
\usepackage{subfigure}
\usepackage{graphicx}

\def\IMPL{\texttt{WAM2EAM}}

\floatstyle{ruled}
\newfloat{Algorithm}{ht}{lop}

\input{boxes}

\def\RULE#1{\vspace{0.2ex} \item[\textsf{#1}]}

\title{Casting of the WAM as an EAM}

\author{%
  Paulo André\inst{1} \and
  Salvador Abreu\inst{1}}
\authorrunning{Paulo Ricardo André}

\institute{
  Departamento de Informática, \\
  Universidade de Évora and CENTRIA FCT/UNL, Portugal \\
  \email{\{prla,spa\}@di.uevora.pt}
}

\begin{document}

\maketitle{}

\begin{abstract}
  Logic programming provides a very high-level view of programming,
  which comes at the cost of some execution efficiency.  Improving
  performance of logic programs is thus one of the holy grails of
  Prolog system implementations and a wide range of approaches have
  historically been taken towards this goal.  Designing computational
  models that both exploit the available parallelism in a given
  application and that try hard to reduce the explored search space
  has been an ongoing line of research for many years.  These goals in
  particular have motivated the design of several computational
  models, one of which is the Extended Andorra Model (EAM). In this
  paper, we present a preliminary specification and implementation of
  the EAM with Implicit Control, the \IMPL{}, which supplies regular
  WAM instructions with an EAM-centered interpretation.
\end{abstract}

\section{Introduction}
\label{sec:introduction}
Logic programming provides an abstract and high-level view of
programming in which programs are expressed as a collection of facts
and predicates that define a model of the problem at hand and against
which questions may be asked.  The most well-known example of this
paradigm of programming is Prolog, which has been sucessfully used in
applications of many different areas.  One line of work that has been
followed to address performance issues is parallel execution:
parallelism allows logic programs to transparently exploit
multi-processor environments while extensions like co-routining,
constraints and tabling go a long way towards reducing the problem's
inherent search space.  Some or all of these together act as the
foundation on which to build more advanced techniques towards
obtaining maximum performance.

From the experience gained in implementing the Basic Andorra Model,
D.H.D.~Warren made a more radical proposal, the Extended Andorra
Model, or EAM~\cite{eam-slides}, in which the conditions in which
independent computations might be carried out are eagerly sought.  In
this article, we present a concrete implementation of the Extended
Andorra Model, the \IMPL{}, which differs from other approaches taken
in the past because we are compiling straight WAM code into
C,\footnote{We are targetting C with GCC extensions, such as label
  values and indirect jumps.}  adopting an EAM computational model,
resorting to GCC extensions.

This paper is structured as follows: Section~\ref{sec:related}
presents a short survey on the road leading up to our current
implementation as far as the EAM is concerned, from the Andorra
Principle to the BEAM.  Section~\ref{sec:eam} describes the EAM in
more detail and lays down the theoretical groundwork of the \IMPL{}
and delves more deeply into its practical
implementation from WAM code compilation to the data structures and
execution control of the EAM-based generated C code.  Finally,
Section~\ref{sec:example} uses a concrete Prolog example and succintly
describes its compilation and execution from start to finish.

\section{State of the Art and Related Work}
\label{sec:related}


A significant body of research on Logic Programming has been directed
towards improving the performance of Prolog.  One important line of
research towards this goal is the exploitation of the different forms
of implicit parallelism, present in Prolog programs.  Several
approaches have been devised over the years but we shall focus on the
systems which allow for the transparent parallel goal execution, in
particular the ``Andorra'' family of languages which includes
Andorra-I, AKL and the BEAM.

\subsection{The Andorra Principle}
\label{sec:andorra-principle}

David H. D. Warren proposed the \textbf{Basic Andorra Model}
(BAM),\footnote{Not to be confused with Van Roy's Berkeley Abstract
  Machine, used in the Aquarius Prolog system~\cite{Aquarius90}.}
geared towards the execution of logic programs, in which a goal is
called \emph{determinate} if it has at most one candidate clause.  In
this model, deterministic goals should be executed first, thereby
reducing the nondeterminate ``guesswork'' to the minimum possible.
Only then, once no deterministic goal remain to be executed, should a
non-deterministic goal be selected for execution.

A system incorporating the Andorra Principle reduces the search space
of logic programs by having deterministic goals execute first and only
once, rather than have them re-executed several times in different
points of the search space.  This behavior is also known as
``sidetracking.''  Also, as a desirable consequence, deterministic
goals may generate constraints (bindings) which may further reduce the
number of alternatives in other (non-deterministic) goals, possibly
even making them deterministic.

Another interesting advantage is how all deterministic goals can
execute in parallel, so long as they do not run into binding
conflicts.  Parallelism in the BAM comes in two flavours:
\begin{itemize}
\item \textit{AND-Parallelism} - deterministic goals run in parallel
\item \textit{OR-Parallelism} - the exploration of different
  alternatives to a goal is done in parallel
\end{itemize}
The BAM may also alter the semantics of programs, in that the order of
the solutions for a given goal may be different from that resulting
from sequential Prolog execution.  This may cause otherwise
nonterminating programs to reach a solution.

There are, however, a few issues inherent to this sort of
computational model:
\begin{itemize}
\item Finding which goals are deterministic can sometimes be difficult
  as predicates with more than one clause may actually have a single
  matching clause for a given query.
\item Concurrency may break Prolog semantics, for instance by
  executing a pruning directive (e.g.~cut) too early.
\end{itemize}
The best-known implementation of the Basic Andorra Model is
Andorra-I~\cite{DBLP:conf/iclp/CostaWY91a,DBLP:conf/ppopp/CostaWY91}.
It exploits OR-parallelism and determinate dependent AND-parallelism
while fully supporting Prolog, however, despite good results, the
system is limited by the fact that co-routining and AND-parallelism
can only be exploited between determinate goals.

Shortly after, Warren went further and proposed the \textbf{Extended
  Andorra Model} (EAM) which improved upon the ideas of the BAM,
namely by trying to explore independent AND-parallelism.  This lead to
a two major approaches:
\begin{itemize}

\item \textbf{AKL}: The Andorra Kernel Language
  (AKL)~\cite{DBLP:conf/elp/FranzenHJ91,DBLP:conf/slp/JansonH91} was
  designed by Haridi and Janson and was the course followed at SICS.
  It concentrated on the idea that a new language was needed, based on
  the advantages of the EAM, which would subsume both Prolog and
  committed-choice languages.  AKL distinguished itself by featuring
  an \emph{explicit} control scheme, as programs were written using
  guarded clauses, where the guard was separated from the body with a
  sequential conjunction, cut or commit operator.

\item \textbf{EAM with Implicit Control}: In contrast to AKL, David
  H. D. Warren and other researchers at Bristol worked towards an
  implementation of the EAM with implicit control.  Its main goal was
  to take advantage of the Andorra Principle while alleviating the
  burden on the programmer.

\end{itemize}

\subsection{The BEAM}
\label{sec:beam}

The Boxed EAM (BEAM) is an implementation of the EAM design with
implicit control, developed at University of Porto,
Portugal~\cite{DBLP:conf/agp/LopesC99,DBLP:conf/padl/LopesCS01,DBLP:conf/epia/LopesCS03,lopes97:beam}.
The beam's initial goal was to prove the feasibility of Warren's
design for the EAM, and as a first step it concentrated on the
original rewriting rules of the EAM, so formally it was defined
through rewrite rules that manipulate AND-OR trees as well as
simplification and optimization rules used to simplify the tree and
discard boxes.  It also made use of a general control strategy, which
is used to decide when and how to apply each rule.

The main operations of the BEAM are:
\begin{itemize}
\item \textbf{Reduction} expands a goal G into and or-box.
\item \textbf{Promotion} promotes constraints from the an inner
  AND-box to an outer AND-box.
\item \textbf{Propagation} propagates constraints from an outer
  AND-box to the inner-boxes.
\item \textbf{Splitting} distributes a conjunction across a
  disjunction.
\end{itemize}
Adding to these are a few simplification and optimization rules, all
of which are described in~\cite{DBLP:conf/epia/LopesCS03}.

Apart from AND- and OR-boxes, there's also another kind of box
contemplated in the BEAM which is the choice-box.  These are special
OR-boxes created when the clauses defining a procedure include a
pruning operator, generically designated by \texttt{\%}.  The original
EAM supports two pruning operators, cut and commit.

The EAM tries to keep the control implicit as much as possible,
contrary to AKL for instance.  Therefore, in the BEAM, the control
decisions are based exclusively on information implicitly extracted
from the program.  Moreover, one of the main goals of the EAM is to
perform the least possible number of reductions to obtain the
solutions to a goal.  BEAM's control strategy is geared towards this
goal.

The BEAM also does not attempt to do all the work by itself, instead
relying on the output of an existing Prolog compiler, in this case YAP
Prolog.  The BEAM was built as an extension to YAP.  It differs from
the work reported herein in that the BEAM is meant to be an
interpreter, whereas \IMPL{} takes WAM code and compiles it to C.

\subsubsection{Non-termination}
\label{sec:non-termination}

A central problem found by the developers of the BEAM was a
consequence of EAM's execution scheme: as long as they do not bind any
(external) variables, the EAM allows the early parallel execution of
nondeterminate goals.  In the worst case, this may lead to
non-termination for certain recursive predicates.  The proposed
solution was based on both eager non-determinate promotion and tabling
which, on the one hand guarantees that the computation ends in
programs that have finite solutions and on the other hand, with
tabling, allows for the reuse of solutions to goals.

\section{The Extended Andorra Model and \IMPL{}}
\label{sec:eam}


The Extended Andorra Model (EAM) is the foundation for the work we
carried out with \IMPL{}.  The ideia is to perform as much work as
possible in parallel, exploiting all the avaliable forms of
parallelism:
\begin{itemize}
\item \textit{Or-parallelism}, related to exploring the various
  alternatives of any given goal.
\item \textit{Indendent AND-parallelism}, within a conjunction of
  goals that do not share any variables.
\item \textit{Dependent AND-parallelism}, between goals that \emph{do}
  share variables.
\end{itemize}
The main extension of the EAM over the BAM is that non-deterministic
goals are allowed to execute in parallel so long as they do not bind
any external variables.




Our purpose is to provide a concrete implementation of the EAM with
implicit control.  It departs from existing work because it compiles
regular WAM code into C, using an EAM runtime specification.
Therefore, the biggest challenge and arguably the most interesting
aspect of this work, is going from one paradigm (Prolog compiled onto
the WAM) to a different one (EAM) with a single tool.




Based on a configuration AND-OR tree at all times, the way to evolve
this configuration is by using one of several \textit{rewrite} rules
on it and an execution control scheme to manage the application of
these rules. %
\ifpaper %
We do not present the rewrite rules used by \IMPL, as these are
closely related to those presented in~\cite{eam-slides}.
\fi %


The major challenge in \IMPL{} certainly is to go from a WAM program
and re-interpret it from an EAM point of view.  To accomplish that, we
take the GNU Prolog's textual WAM output and proceed from there.  The
idea is to generate C code for an EAM runtime. This entails doing
things quite differently from previous work such as WAMCC~\cite{wamcc}
or B-Prolog~\cite{TA98:BIN}.  \IMPL{} has two major aspects to
it:
\begin{enumerate}
\item the compiler, comprising the parser and the C code generator,
\item the runtime, a collection of data structures, logic and
  execution control that implements the EAM execution model.
\end{enumerate}

\noindent The remainder of this section discusses design and
implementation of the compiler and runtime.  We work a running example
based on the sample Prolog code shown in figure~\ref{fig:sample-code}.
\begin{figure}[htbp]
  \centering
\begin{verbatim}
        main :- p(X).

        p(X) :- q(X), r(X).

        q(1).  q(2).
        r(2).  r(3).
\end{verbatim}
  \caption{Sample program}
  \label{fig:sample-code}
\end{figure}

\subsection{Parsing WAM instructions}

We used GNU Prolog because its compilation passes are fairly simple
and it is easy to materialize the WAM representation of Prolog
programs.  The following is a snippet of code which is the GNU Prolog
WAM representation of the \texttt{p/1} predicate from the earlier
example.

\begin{verbatim}
        predicate(p/1,5,static,private,user,[
            allocate(1),
            get_variable(y(0),0),
            put_value(y(0),0),
            call(q/1),
            put_value(y(0),0),
            deallocate,
            execute(r/1)]).
\end{verbatim}


\ifpaper %
\noindent We built a parser for this representation in Bison, which constructs
an abstract parse tree of the WAM program.
\fi %

\subsection{C Code Generation}
\label{sec:c-code-generation}

\ifpaper %
An interesting aspect of \IMPL{} is how we take a sequence of
instructions intended for the regular WAM and directly re-interpret
them in an EAM context, yielding appropriate patterns of target code.
\fi %
Be that as it may, a lot of the WAM instruction set translates as-is
to the EAM representation. Simpler instructions, such as
\texttt{put\_value} for instance, are supposed to do exactly the same
thing in the WAM and in the EAM and the same goes for indexing
instructions like \texttt{switch\_*}.  In a few cases, such as
\texttt{proceed}, \IMPL{} simply disregards the instruction as not
being useful in the EAM setting.

At closer inspection of the WAM instruction set, the major difference
in paradigm impacting the C code generation concerns the instructions
dealing with non-determinism.  Whereas the WAM deals with choice
points, creating and destroying them as needed, the EAM, by doing away
with the WAM's stack-based representation and using an AND-OR tree
based configuration instead, deals with OR-boxes when it comes to
setting up and exploring alternatives.

Once every detail of the original program has a C representation -- an
abstract parse tree -- the idea is to walk through it and emit a bit
of C for each predicate and for every WAM instruction inside it.  For
each internalized predicate, a block of C code is generated, setting
up a new AND-box which contains a suitable number of allocated local
variables,\footnote{The exact number is determined by inspection of
  the WAM code in the body.} binding those variables to its parent
OR-box corresponding predicate arguments and defining each of those
variables' \emph{home} as the very AND-box that is being created.  The
output code is generated by this code in the compiler:

\begin{verbatim}
   emit(8, "a = new_and_box (o, %d, ab_id++);\n", max_var_idx+1);
   for (i = 0; i < n; i++)
     emit(8, "bind (a->locals[%d], o->args[%d]);\n", i, i);
   for (i = 0; i < max_var_idx+1; i++)
     emit(8, "ASREF(a->locals[%d])->home = a;\n", i);
\end{verbatim}

\noindent \texttt{max\_var\_idx} reflects the maximum number of
variables used in this predicate, accounting for possible temporaries
in all of its clauses, potentially a single one if deterministic.
Looking now at the C code for a clause with two local variables, it
might look something like this:

\begin{verbatim}
   a = new_and_box (o, 2);
   bind (a->locals[0], o->args[0]);
   bind (a->locals[1], o->args[1]);
\end{verbatim}

\noindent This allocates a new AND-box with two local variables, as a
child of the current OR-box (whose address is kept in \texttt{o}) and
both of those variables are then immediately bound to whatever are the
first two parent OR-box arguments.  This creates variable chains
across the AND-OR tree, reflecting the same concept found in Prolog
clauses where a newer variable might refer to an older one.

A second pass through the WAM instructions for the clause is needed to
generate code for each actual WAM instruction by traversing the list
built by the parser.

\begin{verbatim}
   while (instrs) {
     print_instr (instrs->head, (*a)->name, n, max_var_idx+1, FALSE)
     instrs = instrs->tail;
   }
\end{verbatim}

\noindent \texttt{print\_instr} then goes through a large
\texttt{switch} instruction that finds the appropriate bit of C code
to emit for each WAM instruction, having the EAM execution scheme in
mind.  WAM instructions, which by now we regard as EAM instructions in
their own right, are roughly divided in three major groups:

\subsubsection{Choice point manipulation} These are the \texttt{try*},
\texttt{retry*} and \texttt{trust*} instructions.  We no longer think
in terms of choice point frames, instead looking at managing
non-determinism by way of OR-boxes.  A predicate with only one clause
consists of an OR-box with a single alternative (and thus a single
descendant AND-box) whereas a non-deterministic predicate (ie. having
more than one clause) is translated as an OR-box with as many children
AND-boxes as there are possible clauses.  A more in-depth description
of how OR-boxes actually deal with alternatives will be given after we
introduce the major data structures used throughout \IMPL{}.  In
practice, an instruction like \texttt{try\_me\_else (L)} (or
\texttt{retry\_me\_else (L)}, for that matter) for predicate
\texttt{q(1)} simply defines the next alternative in the current
OR-box, generating the following bit of C code:
\begin{verbatim}
   o->alt = &&P_q_1_C4;
\end{verbatim}

\subsubsection{Execution Control} The \texttt{call} and
\texttt{execute} instructions are responsible for predicate calling,
in effect jumping to the appropriate place in the code where to start
executing the called predicate. They also need to setup a return
address for when this predicate is done executing.  This is
accomplished by emiting a C label and configuring the current AND-box
continuation to that label, using GCC's label address extension.  With
this, once the called predicate is done, it will \texttt{proceed} to
whatever AND-continuation is available in its AND-box, in effect
returning here and resuming execution.  The difference between
\texttt{call} and \texttt{execute} is precisely what to do after the
called predicate is done with.  Whereas in the former case, it simply
continues executing whatever is left in the current predicate, the
latter means this was the last goal in the current clause and it
should look for a continuation above, in the Prolog execution
chain.  Here's how the \texttt{call} instruction is translated to C:
For example, the pattern of code generated for calling the goal
\texttt{q(X)} in our example is:

\begin{verbatim}
   /* call(q/1) */
   q_enqueue(a->and_conts,&&R1); // setup AND-continuation
   o = new_or_box(a,1);          // create new OR-box
   o->args[0] = a->locals[0];    // preload A registers on the new OR-box

   goto P_q_1;                   // jump to the predicate's code

  R1:                            // return label
   /* further code.. */
\end{verbatim}

\subsubsection{Variable manipulation and unification} This type of
instructions is also handled quite differently within the EAM.  Simple
instructions such as \texttt{put\_value} or \texttt{get\_variable} are
basically the same, but unification needs to be looked at more
carefully, as trying to bind variables which are not local to the
current AND-box leads to suspension of execution and triggers a search
for work, elsewhere in the code.  AND-box suspension and the \IMPL{}
execution scheme will be looked upon in a bit more detail shortly.

\subsection{Generated code structure}
\label{sec:gener-code-struct}

Since we're generating a valid C program, ready to be compiled by GCC,
there's a question of what layout this code will use. One important
constraint is that we must be able to jump back and forth between
different predicates, in order to implement predicate calling and
returning.  Also, we need to jump to random places in the code when
attempting to resume a suspension.  Considering that it is illegal to
use C's \texttt{goto} between different functions,\footnote{We may not
  reenter an existing C stack frame.} generating one C function per
predicate is not an option, no matter how tidy and comfortable that
would be from a structure point of view.

One possible alternative then is to implement the entire program as a
single function and delimiting predicates using unique labels. This
way, jumping from one point in the code to another remains within the
bounds of the one function and correct indentation when emitting the
code will hopefully not make it a burden to look at.  We also must be
careful when jumping to a point of code from out of nowhere, since the
correct environment must be replaced, namely the current AND- and
OR-boxes.  Other than that, all it takes for jumping around the code
is the address to jump to and making good use of GCC's labels as
values extension.

\begin{verbatim}
        int program ()
        {
          /* ... */
          P_p_1: {
              a = new_and_box(o,1);
              /* ... */
              o = new_or_box(a,1);
              goto P_q_1;

          /* ... */
          P_q_1: {
             a = new_and_box(o,1);
             /* ... */
        }
\end{verbatim}

\subsection{Runtime Data Structures}

The runtime half of \IMPL{} is itself broken into two major steps and
these are where we significantly depart from the WAM way of doing
things and completely focus on EAM.  First, executing the C code
previously generated by the compiler will incrementally build the
\textit{configuration}, an AND-OR tree that gets constructed, modified
and pruned as execution of the code proceeds.  The way for this to
happen is by applying in turn the different AND-OR tree rewrite rules.

The most important data structure in \IMPL{} is the AND-OR tree, also
known as the \textit{configuration}.  An AND-OR tree is so called
because it is composed of two kinds of nodes: the AND nodes,
corresponding to Prolog clauses and the OR nodes, consisting of Prolog
goals.  We'll shortly get into more detail on how both these nodes are
structured and how they interact with each other.  For now, it's
important to note that no two nodes, or \textit{boxes}, of the same
type are directly connected in an AND-OR tree, so any path from the
root to any leaf is always made of alternate types of boxes.  A parent
OR-box has AND-box children, each of which has descendent OR-boxes,
and so on.  Moreover, the root is always an OR-box.

\subsubsection{AND-boxes} They represent clauses, so there is one
AND-box in the configuration for every clause in the Prolog source
code.  So, for instance, a non-deterministic predicate having four
different clauses, would consist of four AND-boxes, one for each
clause.  AND-boxes are a lengthy structure in \IMPL{} in that they
play a critical role.  They are home to the clause's local variables,
they need to keep track of their continuations (e.g. where to find the
code for the next goal in the clause once the current goal is done
with) and they also may or may not be suspended at any point in time.
Finally, promotion also impacts AND-boxes directly, so they also have
mechanisms to deal adequately with that.  And, of course, they spawn
(and in turn descend from) OR-boxes corresponding to the reduction of
their body goals.

\subsubsection{OR-boxes} These represent goals and are created
everytime a new goal is executed.  Their primary concern is dealing
with non-determinism by managing goal alternatives, namely holding an
address for the next alternative for the current goal at all times.
They also carry the goal's arguments when the goal gets called in
order to pass them initially to each clause's AND-box as initial
values.  OR-boxes thus spawn an AND-box for each clause they invocate.

\subsection{Suspensions} \label{sec:suspensions}

As we have seen before, caution must be taken when an attempt to bind
a variable is made.  Only in case the variable is local to the current
AND-box will binding be allowed to occur.  Otherwise, the AND-box is
said to be suspended on the offending variable and execution proceeds
elsewhere, namely to the next alternative in the current OR-box.
Execution can only return to this AND-box when certain conditions are
met, namely when the variable becomes local to the current AND-box or
it gets bound from elsewhere.  In the latter case, when the suspension
is resumed, the attempted binding that triggered the suspension in the
first place is retried and it either checks OK or it fails against the
prevailing (earlier) binding.

In order to correctly deal with these situations, we need to wrap
instructions wherein a suspension might occur with some code that
actually checks for ``offending'' binding attempts, namely trying to
bind a non-local variable.  We do this by having every unification
instruction check whether the dereferenced variable is already bound
and if not, whether it is local or external to the current AND-box.
The result of this verification is then returned as a meanigful code
to a wrapping \texttt{CHECK()} macro, which then acts accordingly.
Faced with a unification attempt, the outcome can then be any one of:
\begin{description}
\item[\textbf{BIND\_SUSP}] the variable is not bound yet and it's not
  local to the current AND-box either.  The current AND-box suspends
  on this variable.

\item[\textbf{BIND\_OK}] the variable is not bound and it's local, so
  the binding succeeds.

\item[\textbf{CHECK\_OK}] the variable is bound and its value is the
  same as the one being attempted in the binding, so execution may
  proceed.

\item[\textbf{CHECK\_FAIL}] the variable is bound and its value
  differs with the one being tried. The configuration branch rooted in
  the current AND-box fails and is pruned off the tree.

\end{description}

\noindent
Because of suspensions, for every non-trivial program it's easy to see
that we quickly arrive at what we call a \textit{stuck configuration},
an AND-OR tree where all leaf AND-boxes are suspended.  As we don't
stop execution anytime a box suspends, it is only when no more code is
left to execute that we have a problem.  At this time we try to apply
one of the rewriting rules, in particular giving priority to
determinate rules such as determinate promotion.  By promoting an
inner AND-box into an outer AND-box, the variables inside it are also
promoted which means they become closer to the AND-box where they will
actually be local, eventually allowing for bindings to happen or
suspensions to resume.

\subsection{Deterministic Promotion}

As explained in the previous section, actions (or rules) that
\textit{contract} the configuration are desirable.  On the other hand,
expanding goals also expands the configuration, as AND-boxes give way
to OR-boxes which in turn give way to more AND-boxes and so forth.
Deterministic promotion, being the only rule that eliminates boxes, is
highly sought after.  This rule is only applicable to OR-boxes with a
single alternative.

Implementation-wise, promoting an AND-box context (variables,
suspensions and continuations) into another requires maintaining their
environments coherent.  In other words, if the resulting AND-box
contains the union of both sets of locals variables from the two
AND-boxes involved in the suspension, then what was the first variable
in the inner (promoted) AND-box is probably no longer the first
variable in the outer (resulting) AND-box after promotion.  This lends
itself to all kinds of mayhem when code still refers
\texttt{a->locals[0]} (WAM register \texttt{X(0)}) when the actual
variable is now at \texttt{a->locals[1]}.

\begin{figure}[htbp]
\begin{center}
\includegraphics[scale=0.40]{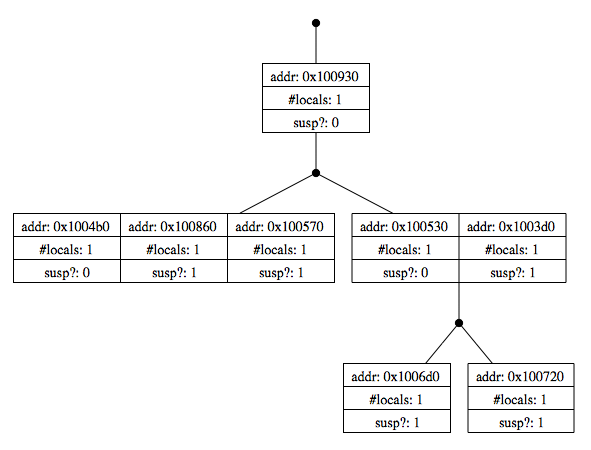}
\end{center}
\caption{On the left: an AND-box grouping made of 3 different AND-boxes.}
\label{fig:and-grouping}
\end{figure}

To cope with this problem, we opted to introduce the concept of
AND-box groupings.  Each AND node in the configuration is actually a
group of one or more complete AND-boxes, forward-connected among
themselves by a pointer which indicates the next box in the group.
Moreover, every box in the group is also linked to the first - the
\textit{head}.  This situation is illustrated in
figure~\ref{fig:and-grouping}.

This way, each box environment remains pristine, as originally
constructed, and it's safe to resume from a suspension point as far as
accessing local variables is concerned.  It's important to note that a
variable is local to the current AND-box if, after dereferencing, its
home AND-box is in the same group, i.e.~has the same head.

\subsection{OR-split and non-deterministic promotion}

Desirable as deterministic promotion might be, its occurence is
heavily constrained as we have shown in the previous section.  The
OR-box must have a single alternative and for predicates with multiple
clauses that's frequently not the case.  It is quite common for a
configuration to get stuck with no chance for deterministic promotions
to occur.  When it comes to this, there is no other choice than to
perform what we call an \emph{OR-split} which forces a situation where
a determinate promotion may happen.

Simply put, we elect an OR-box with more than one alternative to act
as the root of a subtree to be \textit{cloned}.  In the original
subtree, only one alternative remains, while in the cloned subtree,
\textit{every other} alternative is present.  This way, all
alternatives remain in the overall configuration, ensuring correctness
of the program, yet an opportunity for deterministic promotion now
exists.  Note that if the selected OR-box contains only two
alternatives, we arrive at the special case where the OR-split induces
two different deterministic promotion possibilities: one in the
original box and another in the cloned box.

The choice of OR-box to split may be guided by heuristics, yet at this
early stage we're simply going with the leftmost OR-box suitable for
splitting.  Also, from the chosen box's alternatives, we're picking
the leftmost one to remain in the original branch and all others to be
moved to the cloned subtree.  Actual cloning is only needed for the
parent AND-box and any siblings of the chosen OR-box.  OR-split is the
least desirable rule, because with cloning entire branches of the
tree, it quickly becomes expensive.

\subsection{The scheduler}
\label{sec:scheduler}

The need to decide which rule to apply led to the implementation of a
scheduler.  This scheduler is called the first time after all
alternatives and continuations are exhausted and no answers were
produced.  In other words, when the tree is stuck we ask the scheduler
for guidance.

The implementation of the scheduler is part of the runtime code and is
implemented as a C macro.  It basically follows a hierarchy of
possible events and acts accordingly for each outcome.  First of all,
in the event that a variable that had suspensions got bound, it tries
to resume from any suspension pending on that variable.  If none are
found, it looks for an alternative in the current OR-box.  If found,
it continues execution from there, otherwise it tests the tree to see
if it's stuck.  If it is, it tries to apply deterministic promotion in
order to try to move on or, if that fails, it resorts to applying
non-deterministic promotion, by way of an OR-split.  Putting this as
the last choice makes sense, because it is also the most expensive
operation.

\begin{figure}[htbp]
\begin{center}
\includegraphics[scale=0.63]{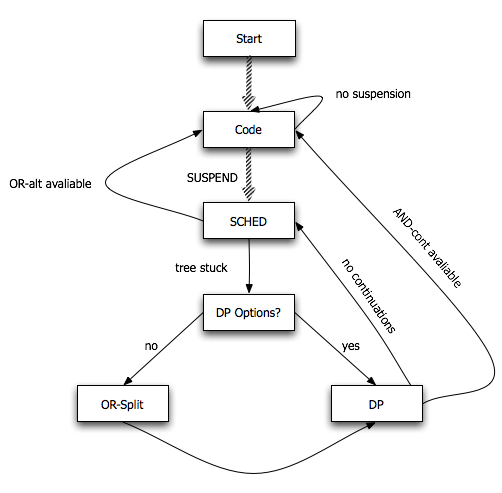}
\caption{The scheduler's flow diagram.}
\label{fig:scheduler}
\end{center}
\end{figure}

It's interesting to note the reason why the scheduler is implemented
as a macro instead of a function, despite being a little involved and
lengthy, it is because it may involve jumping to any point in the
code, be it a suspension point, a continuation or an OR-alternative.
Again, we are faced with the problem of not being able to jump between
different C functions, so its being a macro is sufficient.  The
control flow for the scheduler is depicted in
figure~\ref{fig:scheduler}.

\section{Example Execution}
\label{sec:example}

Getting back to our previous Prolog example of
figure~\ref{fig:sample-code}, we now give an overview of how \IMPL{}
goes from WAM code to EAM execution.

\begin{figure}
\begin{center}
\includegraphics[scale=0.4]{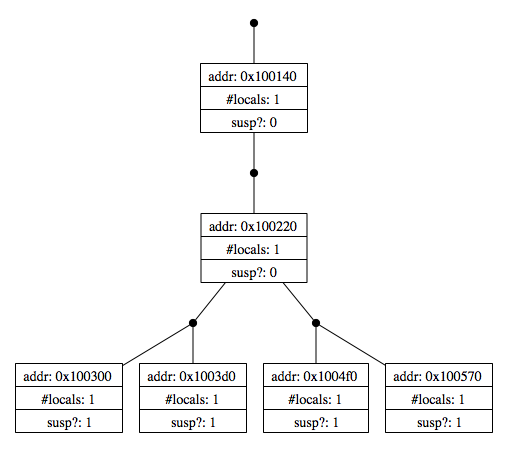}
\caption{Stuck configuration.}
\label{fig:stuck-config}
\end{center}
\end{figure}

As previously mentioned, \IMPL{}-produced C code, when executed,
comprises two diferent phases.  The first one is made of consecutive
reduction steps, expanding the AND-OR configuration as execution
continues through the code.  For each called predicate, a new OR-box
is allocated, spawning a child AND-box for each of the predicate's
clauses.  Only trivial examples won't lead to suspension, as variable
chaining between different predicates immediately induces external
variables on some AND-boxes.  This means that, at first, almost any
binding attempt will lead to AND-box suspension, forcing execution to
look elsewhere in the code, namely in the current OR-box's next
alternative.  So, in our example, it's easy to see how variable X is
only local to the AND-box corresponding to the only clause for the
main predicate and thus every fact for \texttt{q} and \texttt{r} will
lead to suspension over X.  After all clauses are executed, we get to
a \textit{stuck configuration} as seen on
figure~\ref{fig:stuck-config}.

From now on, the configuration is modified by repeated application of
rewriting rules, managed by the \IMPL{} scheduler.  In this case, as
no OR-box contains a single (suspended AND-box) alternative, no
deterministic promotion is possible, so we need to resort to applying
the OR-split rule on the leftmost OR-box, parent to suspended
AND-boxes.

\begin{figure}
\begin{center}
\includegraphics[scale=0.4]{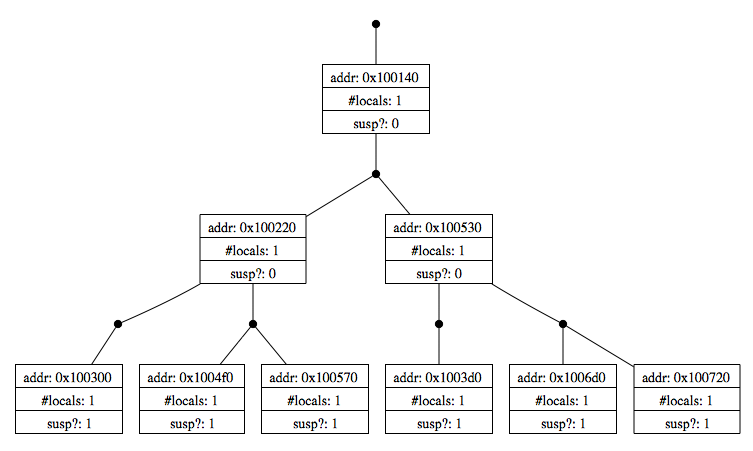}
\caption{After OR-split.}
\label{fig:after-or-split}
\end{center}
\end{figure}

\begin{figure}
\begin{center}
\includegraphics[scale=0.4]{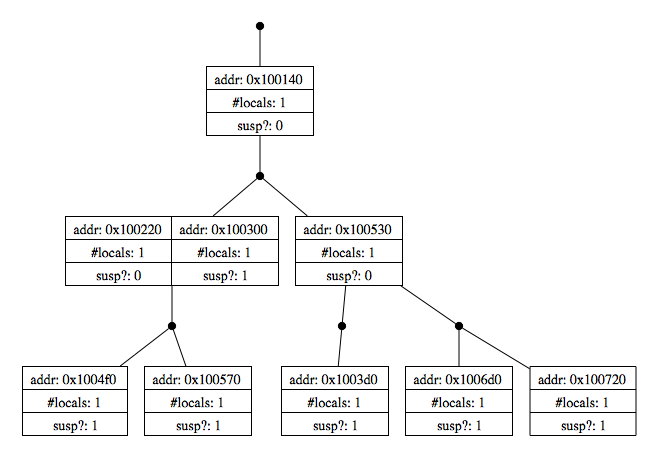}
\caption{After deterministic promotion.}
\label{fig:after-det-prom}
\end{center}
\end{figure}

Because there were only two alternatives to the split OR-box, it means
one stays in the original branch while the other is moved to to the
cloned branch and two determinstic promotions spots now exist
(Figure~\ref{fig:after-or-split}).  Were there more alternatives and
only one deterministic promotion opportunity (in the original branch),
the cloned branch would hold two or more alternatives, and thus not be
ready for deterministic promotion.  So we apply deterministic
promotion to the leftmost AND-box, resulting in the configuration
shown in figure~\ref{fig:after-det-prom}.

After promotion, we attempt to restart the promoted and previously
suspended box, but it immediately suspends again as its local
variable, when dereferenced, still belongs to an another AND-box - the
\texttt{main} AND-box in this example.  Applying promotion to the
analogous case in the cloned branch leads to exactly the same outcome,
so we're again at a stuck configuration scenario.  From here on, it's
easy to see that repeated application of the \IMPL{} rules will result
in a sequence of \textit{OR-split, promotion, suspension} until an
AND-box suspension is restarted and the variable in it that caused
suspension is finally local to the current AND-box group.  When that
happens, the binding succeeds (or fails) and every AND-box suspended
on this variable is ``awakened''.  Then, there are two possible
outcomes:
\begin{itemize}
\item \textbf{CHECK\_OK} -- the attempted binding at the suspension
  point unifies with the one already in place.  This generates an
  answer to the program query and in our example that answer is
  \texttt{X=2}.
\item \textbf{CHECK\_FAIL} -- the attempted binding fails to unify and
  that means this entire branch rooted on the current AND-box group,
  simply \textit{fails} and is pruned from the configuration.
  Execution then looks to the scheduler for where to proceed.
\end{itemize}

As the EAM (and in turn \IMPL{}) does not contemplate explicit
backtracking, the way to generate other answers for any given program,
is to continue exploring different branches of the configuration
looking for other successful bindings.  In this case, none could be
found as the other branch would also have a conflicting binding,
leading to its pruning off the tree.

\section{Concluding Remarks \& Future Work}
\label{sec:conclusion}

We are convinced that our goal of generating a program following EAM
semantics from a classical WAM one has been met, even if with some
restrictions for the time being.  Performance is not yet an issue but
will become one as we develop further aspects of this implementation.
It is interesting to see that it is feasible to have an EAM execution
model without the Prolog compiler being aware of the fact.

Further work is to focus on the introduction of pruning operators --
in the case of cut this is straightforward to recognize from the WAM
code but for \emph{commit} special measures will have to be taken as
it is not inherently accounted for by the Prolog-to-WAM compiler of
GNU Prolog.

One of the driving motivations for generating AND-OR trees and having
them manipulated as per the EAM was to bridge this computational model
to one with tabling, as found in XSB or YAP Prolog.  Although we
haven't begun to do so, this goal remains valid.

There are not many EAM implementations; we need to experimentally
assess our work comparing it to the BEAM and other Prolog
implementations in terms of performance, particularly when we work
towards a parallel version of \IMPL{}.


\section*{Acknowledgments}

The authors would like to thank Ricardo Rocha for fruitful discussions
on the implementation of \IMPL{}.  The FCT (Portuguese Government
Agency) is acknowledged for supporting this work under the project
STAMPA (PTDC/\-EIA/\-67738/\-2006).

\bibliographystyle{plain}
\bibliography{wam2eam,par,lp}

\end{document}